\documentclass{PoS}

\usepackage{amsmath}
\usepackage{amssymb}

\makeatletter
\@ifundefined{showcaptionsetup}{}{%
 \PassOptionsToPackage{caption=false}{subfig}}
\usepackage{subfig}
\makeatother

\title{Lattice Study of the Extent of the Conformal Window in Two-Color Yang-Mills Theory}

\ShortTitle{Lattice Study of the Extent of the Conformal Window in Two-Color Yang-Mills Theory}

\author{\speaker{Gennady Voronov}  (for the LSD Collaboration)%
       \\Department of Physics, Yale University, New Haven, CT 06511, USA\\
       E-mail: \email{gennady.voronov@yale.edu}}

\abstract{
We perform a lattice calculation of the Schr\"odinger functional running
coupling in $SU(2)$ Yang-Mills theory with six massless Wilson fermions
in the fundamental representation. The aim of this work is to determine
whether the above theory has an infrared fixed point. Due to sensitivity
of the $SF$ renormalized coupling to the tuning of the fermion bare
mass  we were unable to reliably extract
the running coupling for stronger bare couplings.}

\FullConference{XXIX International Symposium on Lattice Field Theory \\
                July 10 - 16 2011\\
                Squaw Valley, Lake Tahoe, California}

\begin{document}

\section{Introduction}

A SM Higgs adequately accounts for all electroweak measurements (for
now). However, it has a number of theoretical shortcomings, chiefly
among them the hierarchy problem. Technicolor is a promising alternative
for electroweak symmetry breaking that avoids the introduction of
a fundamental light scalar particle. In particular, a Technicolor
model based on a walking gauge theory can properly account for the
Standard Model fermion masses\cite{appelquist01}. Since such a theory
is expected to reside just below the conformal window, it is essential
to narrow down the extent of this window. The goal of this work is
to do that for the $SU(2)$ gauge theories with $N_{f}$ fermion flavors
in the fundamental representation. These gauge theories are special
in that there is an enhanced $SU(2N_{f})$ global symmetry and it
is currently unknown what implications this will have for technicolor
model building.

The asymptotic properties of a field theory are encoded in the RG
flow of the couplings. We are particularly interested in the infrared
dynamics of two-color asymptotically free theories. Such theories
have two possible distinct IR behaviors. One possibility is that the
$\beta$ function has no zero aside from the one that leads to the
Gaussian UV fixed point. In this case, the running coupling will increase
until it is sufficiently strong to break chiral symmetry. All fermions
will screen out and the coupling will run as in the pure gauge theory.
Alternatively the $\beta$ function may have an additional zero leading
to an IR fixed point. 

There exists a large parameter space of Yang-Mills theories. One can
consider different gauge symmetries and different numbers of fermion
flavors transforming under various representations of the gauge group.
Restricting ourselves to $N_{f}$ flavors of fermions transforming
under the fundamental representation of $SU(N_{c})$ and fixing $N_{c}$,
perturbation theory precisely tells us how many fermions flavors are
required to maintain asymptotic freedom. For two-colors, asymptotic
freedom sets in for $N_{f}<11$. At a fixed color, as we lower the
number of fermion flavors to just below where asymptotic freedom sets
in, we have a weakly-coupled IR fixed point and therefore an IR conformal
theory \cite{Caswell01}. However, for small $N_{f}$ we know that
chiral symmetry is broken in the IR and the theory is confining. Therefore,
we expect that there is some critical number of fermion flavors, $N_{f}^{c}$
at which a theory at fixed color transitions from confining to conformal
IR behavior and we can talk of a conformal window, i.e. the range
of $N_{f}$ at which the theory is both conformal in IR and asymptotically
free in UV. 

The $\beta$ function can be expanded in perturbation theory. The
first two perturbative coefficients are universal and are given in
\cite{Caswell01}. The two loop expansion suggests that $5<N_{f}^{c}<6$
and the six flavor theory has an IRFP around $\bar{g}_{*}^{2}\approx144$.
In perturbation theory, the IRFP at the upper end of the conformal
window is quite weak. As we proceed in decreasing $N_{f}$, the strength
of the perturbative IRFP grows. Therefore, at the lower end of the
conformal window perturbation theory is likely unreliable and we will
need to narrow down the extent of the conformal window using non-perturbative
methods. 

The $SU(2)$ gauge theories with $N_{f}$ flavors of fermions in the
fundamental representation have been previously studied. The $N_{f}=6$
theory is studied by Bursa et. al. They show evidence that the $N_{f}=6$
theory is consistent with an IRFP $5<\bar{g}_{*}^{2}<6$ (in the SF
scheme) \cite{bursa01}. Ohki et. al. show evidence that the $N_{f}=8$
theory has a fixed point $\bar{g}_{*}^{2}>7.5$ using a twisted Wilson/Polyakov
loop method in \cite{ohki01}. It is peculiar that the six flavor
theory as studied by Bursa et. al. would have a weaker fixed point
than the eight flavor theory, especially given how strong the six
flavor perturbative fixed point is. Our goal is to thoroughly study
the $SU(2)$ gauge theories with fermions in the fundamental representation.
We would like to narrow down the extent of the conformal window for
this class of theories. The unexpected results for the six flavor
theory prompted us to begin with it, to see if we could either confirm
or refute the result. Since this work was undertaken, Karavirta et.
al. presented an $SU\left(2\right)$ $SF$ study of the $N_{f}=4$,
6, and 10 theories. They find no evidence of an IRFP, inconclusive
evidence of an IRFP with $\bar{g}_{*}^{2}>11$, and evidence of an
IRFP $\bar{g}_{*}^{2}\approx1$ in the four, six, and ten flavor theories
respectively \cite{karavirta01}.

\section{Schr\"{o}dinger Functional Scheme and Step Scaling}

The Schr\"{o}dinger functional (SF) $\mathcal{Z}$ allows us to define
a non-perturbative renormalized coupling \cite{Luscher02}; it is
given by a path integral over gauge and fermion fields that reside
within a four-dimensional Euclidean box of spatial extent $L$ with
periodic boundary conditions in spatial directions and Dirichlet boundary
conditions in the time direction. We choose gauge boundary conditions
\cite{luscher01}, $\left.U\left(x,\mathrm{k}\right)\right|_{x^{0}=0}=\exp\left[-i\eta\frac{a}{L}\tau_{3}\right]\mbox{ and }\left.U\left(x,\mathrm{k}\right)\right|_{x^{0}=L}=\exp\left[-i\left(\pi-\eta\right)\frac{a}{L}\tau_{3}\right],$
and fermion boundary conditions \cite{sint01}, $\left.P_{+}\psi\right|_{x^{0}=0}=\left.\bar{\psi}P_{-}\right|_{x^{0}=0}=\left.P_{-}\psi\right|_{x^{0}=L}=\left.\bar{\psi}P_{+}\right|_{x^{0}=L}=0.$
The gauge boundary conditions classically induce a constant chromoelectric
background field whose strength is characterized by the dimensionless
parameter $\eta$. With these boundary conditions we see that the
SF $\mathcal{Z}(\eta,L)=\int D\left[U,\psi,\bar{\psi}\right]e^{-S[U,\psi,\bar{\psi};\eta]}.$

The running coupling, in the SF scheme, is defined by, 
\begin{equation}
\frac{k}{\bar{g}^{2}\left(L\right)}=\left.\frac{\partial}{\partial\eta}\log\mathcal{Z}\right|_{\eta=\pi/4}=\left\langle \frac{\partial S}{\partial\eta}\right\rangle ,\label{eq:2.4}
\end{equation}
 where $k=-24\left(L/a\right)^{2}\sin\left[\left(a/L\right)^{2}\left(\pi-2\eta\right)\right]$
is chosen so that the renormalized coupling agrees with the bare coupling
at tree-level. Note that the first two perturbative coefficients of
the SF running coupling are exactly the universal coefficients given
in \cite{Caswell01}. We now have a non-perturbative definition for
a renormalized coupling in a form that is amenable to a lattice calculation.

For this work we used the standard Wilson plaquette gauge action and
the Wilson fermion action. An advantage of working in the SF scheme
is that we can evaluate the running coupling along the $m_{c}\left(g_{0}^{2}\right)$
curve. $m_{c}(g_{0}^{2})$ is defined as the bare mass value that
results in a zero PCAC quark mass \cite{Luchser03}. We determined
$m_{c}$ for a range of bare coupling on $8^{3}\times16$ and $16^{3}\times32$
lattices. We found $m_{c}$ to be consistent within statistics between
the two volumes and consequently interpolated an $m_{c}$ curve in
the $m_{0}-g_{0}^{2}$ plane using the smaller volume data. For $g_{0}^{2}\leq0.5$,
we determine $m_{c}$ using two-loop perturbation theory \cite{Panagopoulos01}.
Otherwise, we use the procedure outlined above. 

We are interested in investigating the running of the coupling over
a large range of scales. A step scaling analysis enables us to do
this in a manner that is computationally feasible \cite{Luscher04}.
We begin by calculating the SF renormalized coupling over a range
of bare couplings and lattice volumes. Lattice perturbation theory
gives $g_{0}^{2}/\bar{g}^{2}$ as an expansion in powers of $g_{0}^{2}$.
This motivates an interpolating fit \cite{appelquiest02}, 
\begin{equation}
\frac{1}{\bar{g}^{2}\left(g_{0}^{2},\frac{L}{a}\right)}-\frac{1}{g_{0}^{2}}=\sum_{i=0}^{n}\sum_{j=0}^{m}a_{i,j}g_{0}^{2i}\left(\frac{a}{L}\right)^{j}.\label{eq:2.5}
\end{equation}
 This procedure produces a smooth function of the renormalized coupling
versus the bare coupling and inverse lattice volume. We note that
it is advantageous to perform one global fit to all data rather than
fitting an interpolating polynomial to each lattice volume. This allows
us to interpolate to additional lattice volumes and smooths out the
approach to the continuum limit. In Figure 2a below we show a bootstrap
replication of our renormalized coupling data plotted against a global
fit.

Now we define the discrete step scaling function, 
\begin{equation}
\Sigma\left(u,\frac{a}{L}\right)\equiv\left.\bar{g}^{2}\left(g_{0_{*}}^{2},\frac{sL}{a}\right)\right|_{\bar{g}^{2}\left(g_{0_{*}}^{2},\frac{L}{a}\right)=u},\label{eq:2.6}
\end{equation}
 it is the value of the renormalized coupling on a volume of $(sL)^{4}$
and bare coupling tuned such that we have a renormalized coupling
of $u$ on a lattice of volume $L^{4}$. We arrive at a continuum
step scaling function, 
\begin{equation}
\sigma\left(u\right)=\underset{\frac{a}{L}\rightarrow0}{\lim}\Sigma\left(u,\frac{a}{L}\right),\label{eq:2.7}
\end{equation}
 by taking the continuum limit of the discrete step scaling function.
In practice, these functions are evaluated using the interpolating
polynomial specified in Eq. (\ref{eq:2.5}) and with the choice $s=2$.
We take the continuum limit by evaluating $\Sigma$ for various values
of $a/L$, taking care to only to use interpolated values of $a/L$,
and extrapolating to zero.

\section{Results}

\begin{figure}
\begin{centering}
\subfloat[]{\includegraphics[height=6cm]{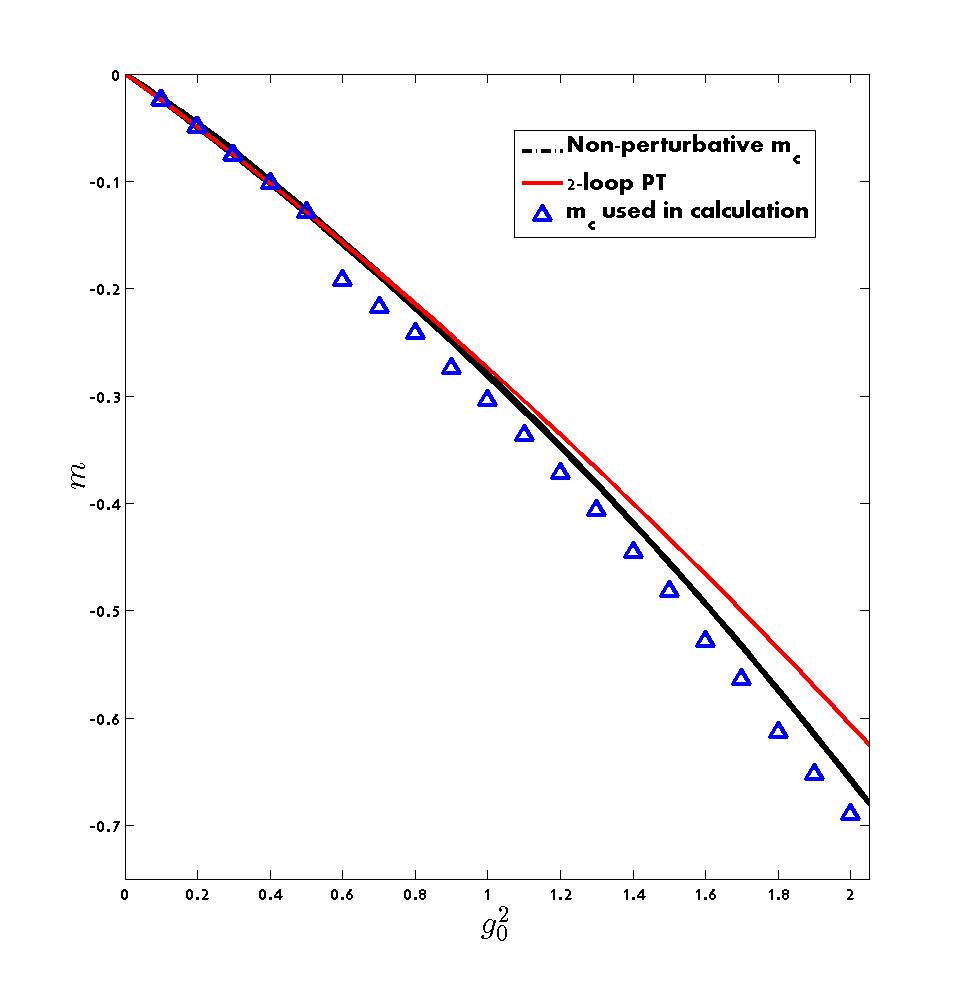}

} \subfloat[]{\includegraphics[height=6cm]{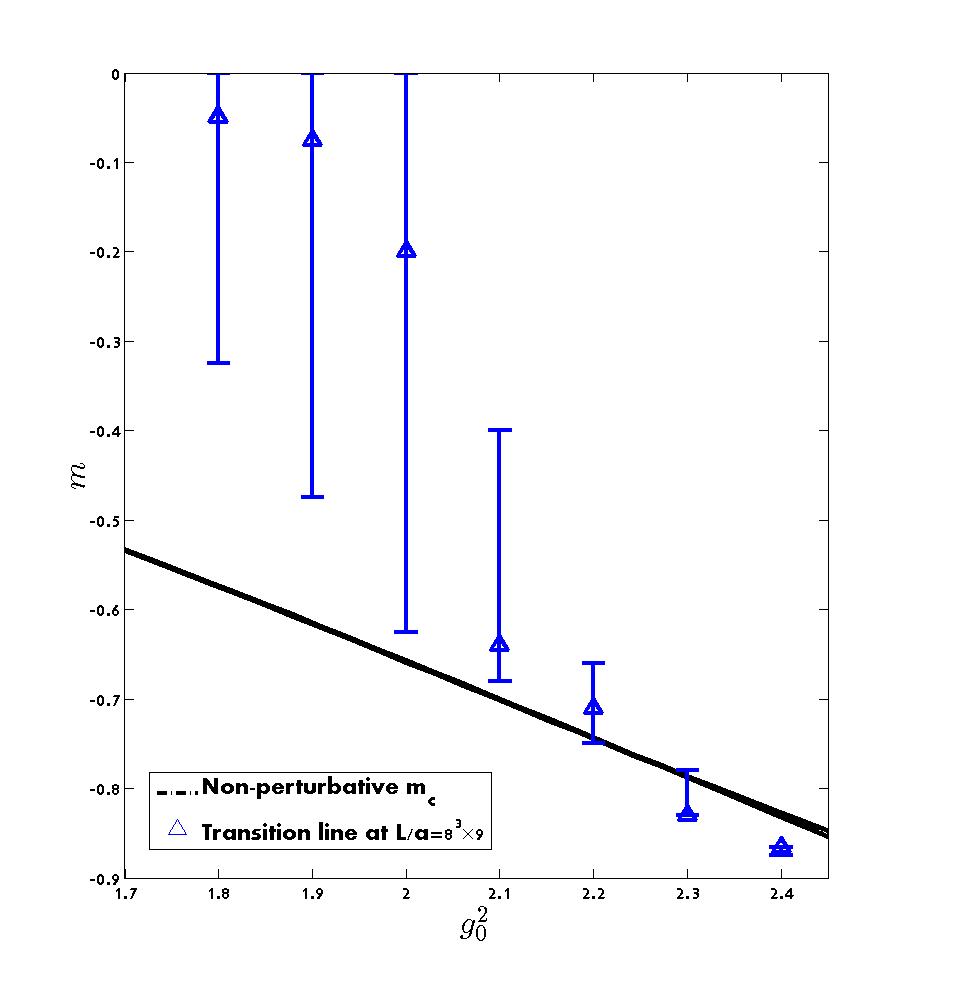}

}
\par\end{centering}

\caption{Critical bare mass to reproduce massless fermions. In (a) continuum
extrapolation of $m_{c}$, 2-loop perturbative $m_{c}$, and values
used in calculation shown. In (b) continuum $m_{c}$ shown against
location of bulk phase transition. }
\end{figure}
 
\begin{figure}
\centering{}\subfloat[]{\begin{centering}
\includegraphics[height=6cm]{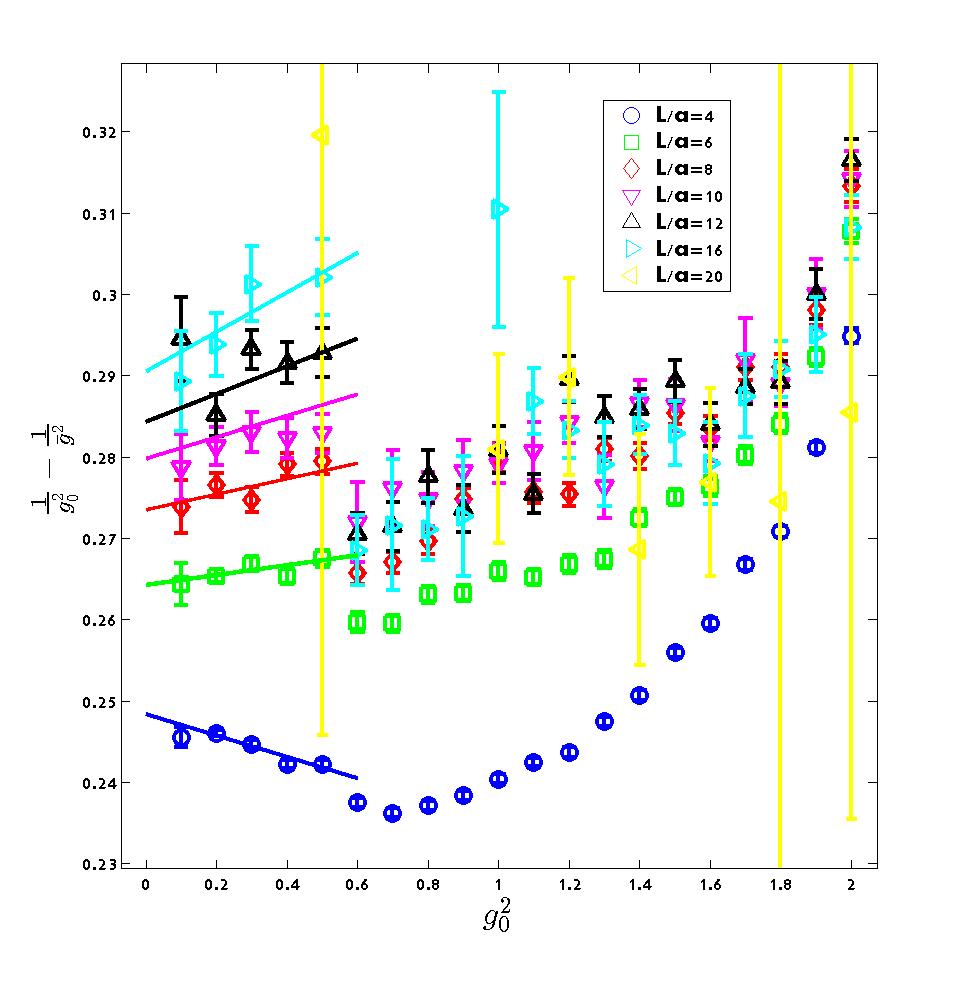}
\par\end{centering}

} \subfloat[]{\begin{centering}
\includegraphics[height=6cm]{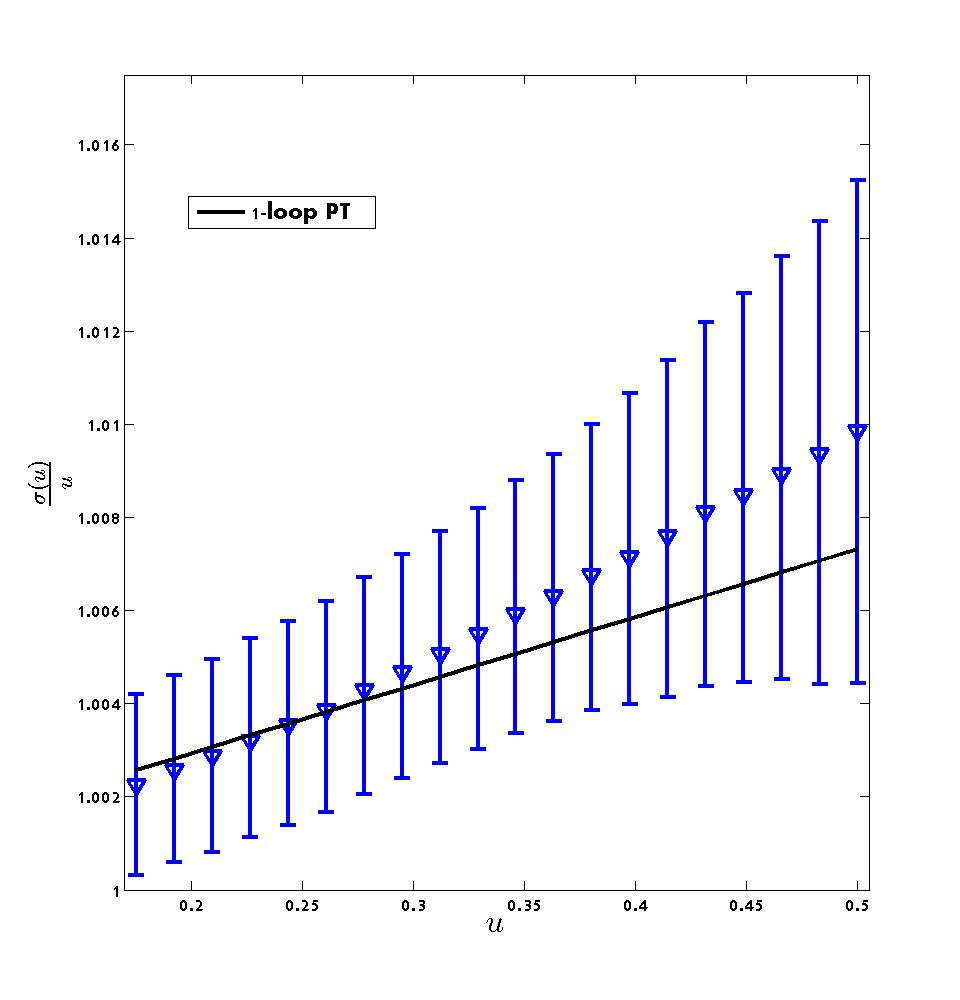}
\par\end{centering}

}\caption{In (a), all renormalized coupling data points in a typical bootstrap
ensemble plotted in form $\frac{1}{g_{0}^{2}}-\frac{1}{\bar{g}^{2}}$
against a global interpolating fit to data restricted to $g_{0}^{2}\leq0.5$.
In (b), a plot of the discrete beta function derived from fit shown
in (a).}
\end{figure}
 
\begin{figure}
\begin{centering}
\subfloat[]{\begin{centering}
\includegraphics[height=6cm]{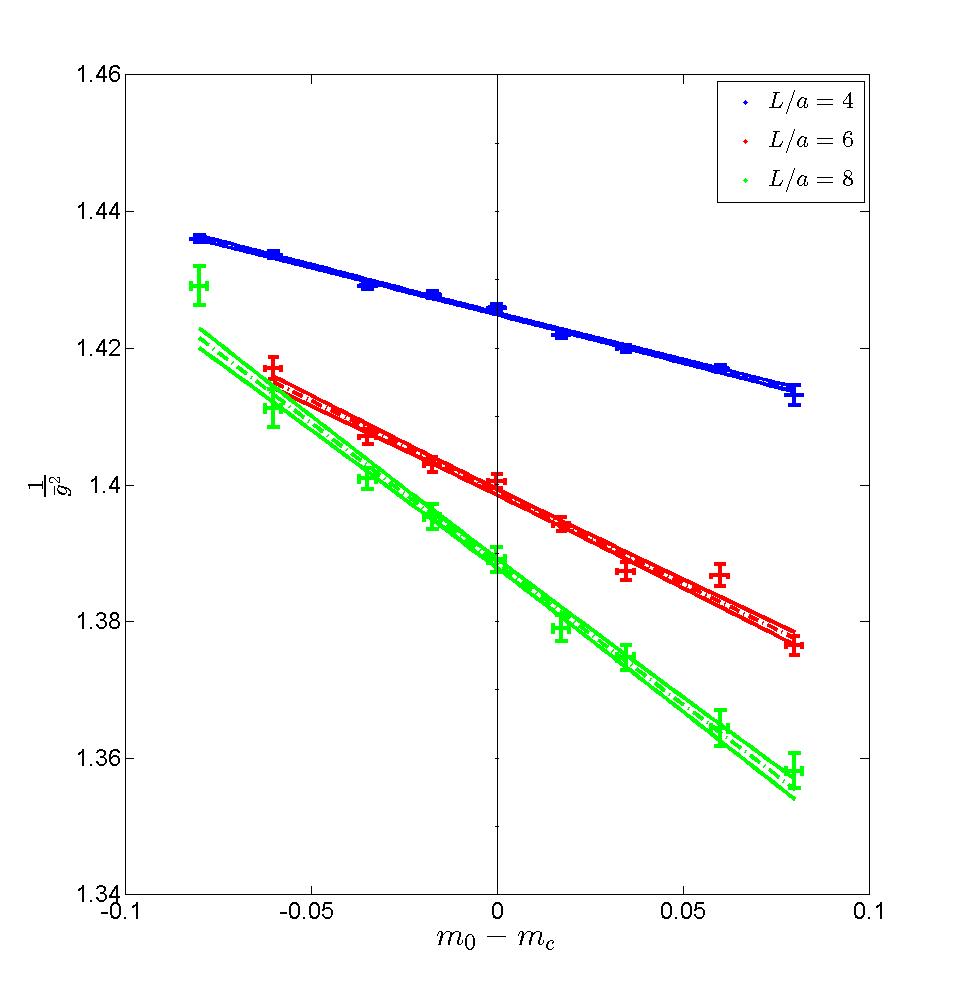}
\par\end{centering}

} \subfloat[]{\begin{centering}
\includegraphics[height=6cm]{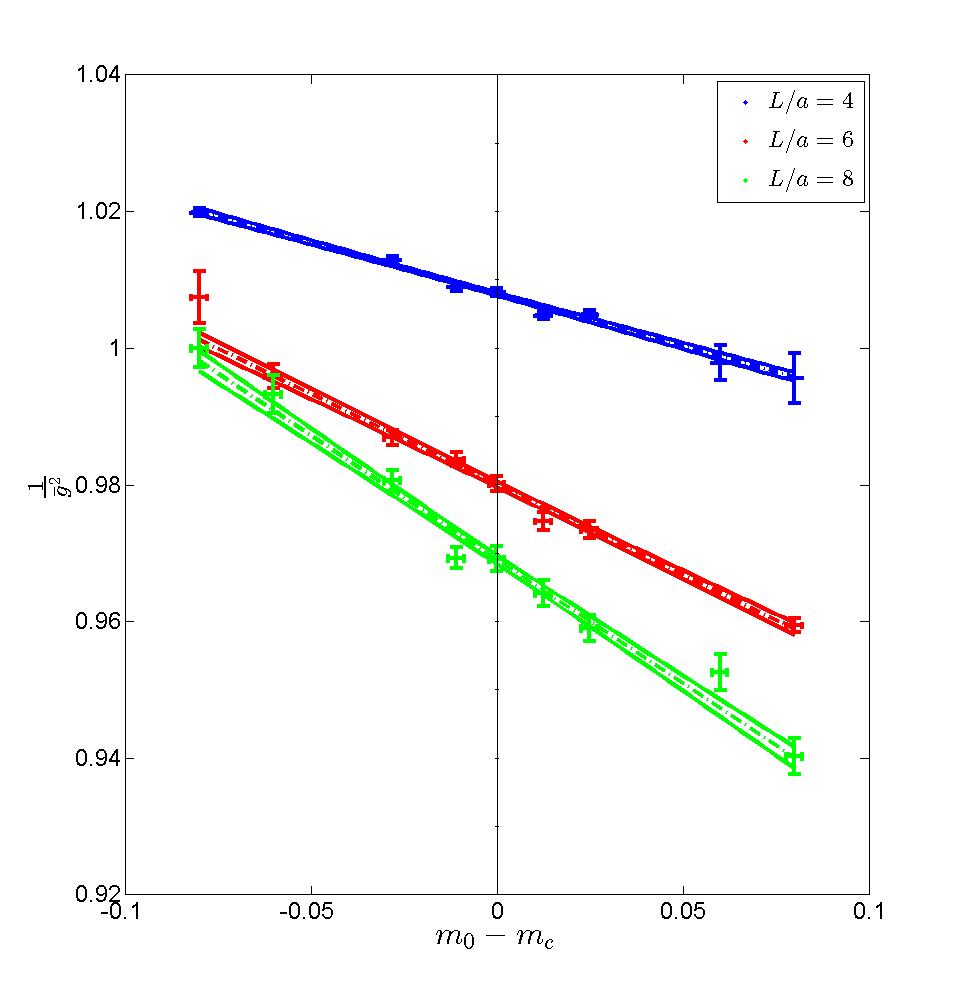}
\par\end{centering}

}
\par\end{centering}

\caption{$\bar{g}^{2}$ evaluated at $g_{0}^{2}=0.6$ and $0.8$, in (a) and
(b) respectively, for various lattice volumes and fermion bare masses. }

\end{figure}

We calculated the renormalized coupling for a range of values of the
bare coupling $g_{0}^{2}$ and lattice volumes $L/a$. Configurations
were generated with the Chroma implementation of the HMC algorithm
and $\frac{\partial S}{\partial\eta}$ is evaluated on every configuration
\cite{chroma}. We aimed to acquire on the order of $100\mbox{ K}$
configurations since $\frac{\partial S}{\partial\eta}$ is a noisy
observable with long autocorrelation times. 

Upon studying the mass tuning on additional lattice volumes and obtaining
additional statistics, we see that our initial observation of no statistically
significant dependence on the lattice volume was incorrect. We calculated
an improved $m_{c}$ curve by fitting our non-perturbatively determined
critical bare mass values to an interpolating polynomial $m_{c}\left(g_{0}^{2},\frac{a}{L}\right)$
and extrapolating to $a/L\rightarrow0$. In Figure 1a, we show the
the values of $m_{c}$ used in our simulations, the two-loop perturbative
$m_{c}$ curve, and our improved non-perturbative $m_{c}$ curve.
It is clear that the values of $m_{c}$ used in our evaluations of
the renormalized coupling deviate significantly from the appropriate
values. Moreover, Figure 2a, which shows our renormalized coupling
measurements versus the bare coupling, indicates a jump discontinuity
at precisely the value of the bare coupling where we switch from using
the perturbatively determined $m_{c}$ to our miss-tuned values of
$m_{c}$. We performed a global fit to all of our renormalized coupling
data that was calculated with a perturbative value of $m_{c}$, i.e.
for $g_{0}^{2}\leq0.5$. The discontinuity in $\bar{g}^{2}$, is highlighted
by extrapolating our fit to the first data points which are mistuned
in the bare mass. To quantify this we evaluated a limited sample of
renormalized couplings at a variety of bare masses near $m_{c}$.
Figure 3 shows $1/\bar{g}^{2}$ versus $m_{0}$ for various lattice
volumes and $g_{0}^{2}=0.6$ and $0.8$. This figure demonstrates
that $\bar{g}^{2}$ is trending in the correct direction to remove
the discontinuity in Figure 2a and that a mistuning of the bare mass
by as little as $4\%$ results in an error in the renormalized coupling
that dominates all other sources of error. Moreover, such an error
can easily introduce or obscure an IRFP, especially if the the RG
flow of the running coupling is particularly slow. 

In order to guarantee that we can take a continuum limit, we need
to ensure that we obtain data from the weak-coupling side of any spurious
lattice phase transition. With this in mind, we scanned through the
bare parameter space and located peaks in the plaquette susceptibility
on a $L/a=8^{3}\times9$ lattice. This search indicates a line in
the $m_{0}-g_{0}^{2}$ plane of first order phase transition that
ends at a critical point at around $g_{0}^{2}\approx2.2$. For $g_{0}^{2}\lesssim2.2$,
we see crossover behavior. In Figure 1b, we show the above transition
line plotted along with $m_{c}(g_{0}^{2})$. Figure 1b indicates that
the six flavor massless fundamental Wilson fermion action has a sensible
continuum limit only for $g_{0}^{2}\lesssim2.2$. Therefore, we expect
that we will not be able to examine the running coupling at sufficiently
strong bare coupling with our current action. 

Finally, using only data generated with a properly tuned bare mass
($g_{0}^{2}\leq0.5$), we study the continuum running of this theory.
Towards this end, we generated $2000$ bootstrap ensembles from our
restricted data set and apply the fitting procedure described in the
previous section to each individual ensemble. A step scaling analysis,
using fits like the one shown in Fig 2a, with $s=2$, using a linear
extrapolation to the continuum using linearly spaced values between
$L/a=$7 and 10, was then used to produce curves, for each individual
ensemble, of the discrete beta function $\sigma\left(u\right)/u$.
Figure 2b shows $\sigma\left(u\right)/u$ plotted against one-loop
perturbation theory. Each point, with two-sided errorbars, was obtained
via the $\mathrm{BC_{a}}$ method \cite{efron01}. This plot demonstrates,
that our methods can properly reproduce perturbation theory when the
bare mass is tuned appropriately.

\section{Conclusions and Outlook }

We were unable to extract a renormalized coupling flow outside the
perturbative region due to mistuning of the bare fermion mass. We
have since properly tuned the bare mass. Figure 2b demonstrates that
our production and analysis applications can reproduce perturbation
theory in a robust manner. We found that these figures do not qualitatively
change as we change the fit parameters and details of the continuum
extrapolation. 

The RG flow of the running coupling in the six-flavor $SU\left(2\right)$
theory is particularly slow and consequently the renormalized coupling
sensitively depends on the bare fermion mass. We emphasize that in
such theories the bare mass must be tuned very carefully. We have
since done this for the $SU\left(2\right)$ six-flavor theory. In
order to reliably study the running coupling of this theory we would
simply have to reproduce our data using the correct value of the fermion
bare mass. 

Studies of the plaquette and plaquette susceptibility indicate the
presence of bulk phase transition along the $m_{c}$ line at $g_{0}^{2}\approx2.2$.
This effectively limits how strong of a renormalized coupling that
we can investigate with the standard Wilson fermion action. Other
investigations into the six-flavor theory suggest that if there exists
an IRFP then it likely resides at a stronger coupling that can be
probed with our current action before encountering a bulk phase transition \cite{karavirta01}.
Therefore, while we can regenerate a new data set with an improved
estimate of $m_{c}$, we believe that it would be a misallocation
of resources to pursue this theory without first switching to an improved
action where the bulk phase transition occurs at a stronger coupling.

\end{document}